\documentclass[pre,aps,twocolumn,groupedaddress,noeprint]{revtex4-2}
\usepackage{amsmath,amssymb,amsfonts}
\usepackage{xcolor}
\usepackage{graphicx}
\usepackage[colorlinks=true,citecolor=blue,linkcolor=blue,urlcolor=blue]{hyperref}

\graphicspath{{./figs}}
\newcommand{\includetikz}[1]{\,\vcenter{\hbox{\includegraphics{diagram_#1.pdf}}}}
\newcommand{\Tc}{T_\text{c}}

\begin{document}

\title{Tensor Renormalization Group Calculations of Partition-Function Ratios}

\author{Satoshi Morita}
\email[]{smorita@keio.jp}
\affiliation{Graduate School of Science and Technology, Keio University,
  Yokohama, Kanagawa 223-8522, Japan}
\affiliation{Keio University Sustainable Quantum Artificial Intelligence Center (KSQAIC),
  Keio University, Minato-ku, Tokyo 108-8345, Japan}

\author{Naoki Kawashima}
\affiliation{Institute for Solid State Physics, The University of Tokyo,
  Kashiwa, Chiba 277-8581, Japan}
\affiliation{Trans-scale Quantum Science Institute, The University of Tokyo,
  Bunkyo-ku, Tokyo 113-0033, Japan}


\begin{abstract}
The behavior of dimensionless quantities defined as ratios of partition functions is analyzed to investigate phase transitions and critical phenomena.
At criticality, the universal values of these ratios can be predicted from conformal field theory (CFT) through the modular-invariant partition functions on a torus. We perform numerical calculations using the bond-weighted tensor renormalization group for three two-dimensional models belonging to different universality classes: the Ising model, the three-state Potts model, and the four-state Potts model. The partition-function ratios obey the same finite-size scaling form as the Binder parameter, and their critical values agree well with the universal values predicted by CFT. In the four-state Potts model, we observe logarithmic corrections in the system-size dependence of these ratios.
\end{abstract}

\maketitle

\section{Introduction}

Phase transitions and critical phenomena are important topics in statistical physics.
Determining the critical temperature and critical exponents is crucial in elucidating the nature of continuous phase transitions.
The finite-size scaling method is a powerful tool for this purpose.
In particular, dimensionless quantities such as the Binder parameter are useful because they require fewer estimated parameters.

The Binder parameter is defined as the ratio of the fourth moment to the square of the second moment of the order parameter~\cite{binder1981critical,binder1981finite}.
Its value in the thermodynamic limit corresponds to the phases, allowing for precise estimation of the critical temperature from the position of its jump.
Moreover, its value for finite sizes follows a finite-size scaling law that depends only on the ratio of the correlation length to the linear size of the system near the critical point.
The critical exponent related to the correlation length can be estimated from the optimal scaling plot of numerical data.
Since the calculation of higher moments of the order parameter is easily performed in Monte Carlo simulations, the Binder parameter has been widely used in many numerical studies.

Recently, numerical methods using tensor networks have been expected to be applicable to systems where existing methods such as Monte Carlo methods are difficult to compute.
The partition function of a lattice model can be represented as a tensor network and can be computed using methods such as the tensor renormalization group (TRG) method and its variants~\cite{levin2007tensor,xie2012coarsegraining,adachi2022bondweighted}.
We proposed a multi-impurity method to calculate higher moments of physical quantities using the higher-order TRG (HOTRG) method~\cite{morita2019calculation}.
Using this method, we calculated the Binder parameter for large systems that are difficult to reach with Monte Carlo methods and estimated the critical temperature and the critical exponent with high accuracy.
We also proposed a multi-impurity method for the bond-weighted TRG (BWTRG) method and obtained even more accurate results~\cite{morita2025multiimpurity}.

The ratios of partition functions proposed by Gu and Wen~\cite{gu2009tensorentanglementfiltering} are dimensionless quantities that can be easily calculated using tensor network methods in a two-dimensional system.
As will be defined later, they are denoted as $X_1$ and $X_2$.
In the thermodynamic limit, they take the same values as the degeneracy of the scale-invariant tensor and can be used to detect phase transitions.
We found in a previous paper that $X_1$ follows the same finite-size scaling as the Binder parameter~\cite{morita2025multiimpurity}.
We also showed that its universal value at criticality can be calculated from the scaling dimensions.

In this paper, we extend these results and analyze the behavior of another dimensionless quantity, $X_2$, and dimensionless quantities defined from larger cluster sizes.
The universal values of them at criticality are derived from the modular invariant partition functions of conformal field theory (CFT).
We perform numerical calculations using BWTRG in three two-dimensional models belonging to different universality classes: the Ising model, the three-state Potts model, and the four-state Potts model.

In the next section, we review the definition of the partition-function ratios and derive their universal values at criticality predicted from CFT.
In Sec.~\ref{sec:numerical_results}, we present numerical results for the Ising model, the three-state Potts model, and the four-state Potts model.
We show the size dependence of the partition-function ratios near the critical temperature and compare them with the universal values predicted from CFT.
In the four-state Potts model, we observe that the partition-function ratios have logarithmic correction with respect to the system size.
In Sec.~\ref{sec:anisotropic}, we consider anisotropic systems and confirm that the universal values of the partition-function ratios depend on the correlation-length ratio as predicted from CFT.
The last section is devoted to the summary and discussion.
We focus on the square lattice models in the main text and the results on the honeycomb lattice are summarized in Appendix.

\section{Conformal Field Theory}
\label{sec:CFT}

We first consider a ratio of the partition functions proposed by Gu and Wen~\cite{gu2009tensorentanglementfiltering} as
\begin{equation}
  X_1(T, L) \equiv \frac{Z_{L\times L}(T)^2}
    {Z_{2L\times L}(T)},
  \label{eq:X1_definition}
\end{equation}
where $Z_{L_1\times L_2}(T)$ is the partition function of a system with dimensions $L_1 \times L_2$ under the periodic boundary condition at temperature $T$.
When an $L \times L$ system is represented by a four-legged renormalized tensor $\mathsf{T}$, $X_1$ can be expressed as
\begin{equation}
  X_1(T, L) = \frac{
    \left(
    \sum_{r, u} \mathsf{T}_{ruru}
    \right)^2
  }{
    \sum_{r,u,l,\tilde{u}} \mathsf{T}_{rulu} \mathsf{T}_{l\tilde{u}r\tilde{u}}
  }
  =\frac{\left(\includetikz{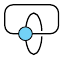}\right)^2}{\includetikz{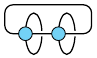}},
  \label{eq:X1_tensor}
\end{equation}
where the indices of $\mathsf{T}$ are arranged in counterclockwise order from the right leg.
In this section and the next, we assume that the system is isotropic, and the anisotropic case will be discussed in Sec.~\ref{sec:anisotropic}.

If we performed a proper renormalization group on the tensor network, we would obtain a scale-invariant tensor corresponding to the phase.
In the disordered phase, the renormalized tensor converges to a trivial tensor, resulting in $X_1 = 1$.
In contrast, in the ordered phase with spontaneous symmetry breaking, the renormalized tensor converges to a non-trivial tensor reflecting the symmetry, leading to $X_1 > 1$.
Therefore, the phase transition can be detected from the jump of $X_1$.
On the other hand, it is known that the naive TRG retains a fictitious fixed-point tensor called the corner double-line (CDL) structure~\cite{gu2009tensorentanglementfiltering}, and thus the above argument may not hold.
However, it would be natural to hypothesise that the tensor obtained through TRG is a tensor product of the scale-invariant tensor and the CDL tensor, i.e.,
\begin{equation}
  \mathsf{T} = \mathsf{T}_\text{inv} \otimes \mathsf{T}_\text{CDL}
  = \includetikz{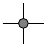} \otimes \includetikz{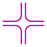}.
  \label{eq:scale_CDL_hypothesis}
\end{equation}
From this hypothesis, the numerator and denominator in Eq.~\eqref{eq:X1_tensor} can be separated into the scale-invariant part and the CDL part as
\begin{equation}
  \begin{aligned}
    \includetikz{01} &= \includetikz{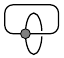} \times \includetikz{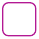},\\
    \includetikz{02} &= \includetikz{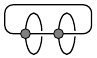} \times \left(\includetikz{17}\right)^2.
  \end{aligned}
\end{equation}
The CDL component cancels out in $X_1$, and the scale-invariant tensor determines the value of $X_1$. The same argument applies to the similar ratios discussed below.
In actual calculations, such a clear separation may not be achieved due to the finite bond dimension.
However, when the correlation length and system size are sufficiently smaller than the length scale limited by the finite bond dimension, accuracy of the TRG calculation of the partition function is high, and thus, the value of $X_1$ is close to the true value.
By searching for the jump position of $X_1$ while keeping the bond dimension fixed, the effective transition temperature $\Tc(\chi)$ under the finite bond dimension $\chi$ can be precisely estimated.

Since $X_1$ is a dimensionless quantity, its finite-size scaling formula near the critical temperature $\Tc$ is expected to be the same as the Binder parameter as
\begin{equation}
  X_1(T, L) = \tilde{f}(L/\xi(T))
  = f(L^{1/\nu} t),
  \label{eq:X1_scaling}
\end{equation}
where $t \equiv (T - \Tc)/\Tc$ is the deviation from the critical temperature, $\xi(T) \sim t^{-\nu}$ is the correlation length, and $\nu$ is the critical exponent of the correlation length.
When we perform the finite-size scaling analysis based on this scaling form using the results of the TRG calculation with the fixed bond dimension $\chi$, the estimated critical temperature should agree with $\Tc(\chi)$ which obtained from the jump position of $X_1$.

At criticality, $X_1$ takes a universal value only depending on the universality class.
This value should be determined by the unstable fixed point corresponding to the critical state and can be predicted from conformal field theory (CFT)~\cite{difrancesco1997conformal,henkel1999conformal,ginsparg1988applied}.
The partition function on a torus in CFT is written as
\begin{equation}
  Z_\text{PP}(\tau) = \sum_{\Delta,\overline{\Delta}} N_{\Delta,\overline{\Delta}}
  \chi_{\Delta}(q) \overline{\chi}_{\overline{\Delta}}(\overline{q}),
\end{equation}
where $\tau$ is the modular parameter and $q \equiv e^{2\pi i \tau}$.
The Virasoro character of the holomorphic part is defined as
\begin{equation}
  \chi_{\Delta}(q)
  \equiv \text{Tr}_{\Delta} \ q^{L_0-c/24}
  = q^{\Delta - c/24} \sum_{n=0}^\infty d_{\Delta}(n) q^n,
  \label{eq:Virasoro_character}
\end{equation}
where $c$ is the central charge and $L_0$ is the zero-mode generator of the Virasoro algebra.
$\Delta$ and $\overline{\Delta}$ are the conformal weights of the primary fields, and $d_{\Delta}(n)$ is the number of independent states at level $n$ in the conformal tower.
The coefficient $N_{\Delta,\overline{\Delta}}$ is a non-negative integer that represents the multiplicity of the primary fields.
The subscript ``PP'' indicates that the periodic boundary condition is applied in both directions.
The partition function on a torus should be invariant under the modular transformations
\begin{equation}
  \begin{aligned}
    S& : \tau \to -1/\tau, \\
    T& : \tau \to \tau + 1,
  \end{aligned}
  \label{eq:modular_transformation}
\end{equation}
which generate the modular group.
This modular invariance imposes strong constraints on $N_{\Delta,\overline{\Delta}}$ and operator content~\cite{cardy1986operator,cardy1986effect}.
Specific examples of the modular invariant partition function will be shown in the next section.

The universal value of $X_1$ can be expressed using the modular invariant partition function as
\begin{equation}
  X_1 = \frac{Z_\text{PP}(i)^2}{Z_\text{PP}(i/2)}
  = \frac{Z_\text{PP}(i)^2}{Z_\text{PP}(2i)},
  \label{eq:X1_CFT}
\end{equation}
where the modular parameters $\tau = i$ and $\tau = i/2$ correspond to the aspect ratios of $1:1$ and $2:1$, respectively.
The second equality follows from the modular transformation $S$.
The universal value of $X_1$ depends only on the scaling dimensions of the scaling operators:
\begin{equation}
  X_1
  = \frac{\left(\sum_{\alpha} e^{-2\pi x_{\alpha}}\right)^2}
  {\sum_{\alpha} e^{-4\pi x_{\alpha}}},
  \label{eq:X1_scaling_dimension}
\end{equation}
where $x_{\alpha} \equiv \Delta_{\alpha} + \overline{\Delta}_{\alpha}$ is the scaling dimension of the scaling operator $\alpha$.
In this expression, $\alpha$ runs over all scaling operators, that is, the primary operators and their descendants.

According to the hypothesis~\eqref{eq:scale_CDL_hypothesis}, the universal value of $X_1$ indicates that
the scale-invariant tensor at criticality is determined by the operator content of CFT, that is,
\begin{equation}
  \mathsf{T}_\text{inv} = \mathsf{T}_\text{CFT}.
\end{equation}
This CFT tensor has infinite bond dimensions and includes not only the primary operators but also all their descendants.
Its tensor trace should give the modular-invariant partition function on a torus as
\begin{equation}
  \includetikz{15} = Z_\text{PP}(i), \quad
  \includetikz{16} = Z_\text{PP}(i/2),
\end{equation}
which is consistent with the Gu-Wen's method for extracting the central charge and scale dimensions from the scale-invariant tensor~\cite{gu2009tensorentanglementfiltering}.
The CFT tensor is associated with the universal singular part of the free energy.
Conversely, the CDL structure is regarded as determining the regular part that depends on the microscopic details of a model.

Gu and Wen also proposed another dimensionless quantity in Ref.~\cite{gu2009tensorentanglementfiltering} as
\begin{equation}
  X_2(T, L) = \frac{
    \left(
    \sum_{r, u} \mathsf{T}_{ruru}
    \right)^2
  }{
    \sum_{r,u,l,\tilde{u}} \mathsf{T}_{rul\tilde{u}} \mathsf{T}_{l\tilde{u}ru}
  }
  = \frac{\left(\includetikz{01}\right)^2}{\includetikz{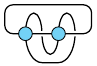}}.
  \label{eq:X2_definition}
\end{equation}
The denominator corresponds to a skewed $2 \times 1$ system, where moving $L$ along the short side causes a shift of  $L$ along the long side.
This system can be represented as a parallelogram with vertices at $(0,0)$, $(2L,0)$, $(L,L)$, and $(3L, L)$.

This skewed boundary condition is geometrically equivalent to a periodic boundary condition of a $\sqrt{2}\times\sqrt{2}$ square tilted at 45 degrees.
This fact can be understood from the modular group.
The skewed boundary condition corresponds to the modular parameter $\tau = (1 + i) / 2$.
Successive application of the modular transformations $S$ and $T$ transforms it to $\tau = i$.
From the point of view of tensor networks, application of TRG to the denominator of Eq.~\eqref{eq:X2_definition} transforms it as,
\begin{equation}
  \includetikz{03}
  \simeq \includetikz{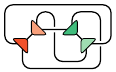}
  = \includetikz{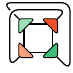}.
\end{equation}
The last diagram is equivalent to the trace of a four-legged tensor tilted at 45 degrees with a scale factor of $\sqrt{2}$.
If a four-legged tensor is scale-invariant, it is equal to the trace in the numerator.

Therefore, the universal value of $X_2$ at criticality is given by
\begin{equation}
  X_2 = \frac{Z_\text{PP}(i)^2}{Z_\text{PP}(\frac{1+i}{2})}
  = Z_\text{PP}(i).
\end{equation}
In the second equality, the denominator of $X_2$ cancels out with a part of the numerator as described above.
That is, the universal value of $X_2$ is the modular invariant partition function itself with a square aspect ratio.

The above discussion can be generalized by considering larger cluster sizes.
For example, from a $3\times 1$ cluster, we can define dimensionless quantities as
\begin{equation}
  X_{3\times 1}(T, L)
  \equiv \frac{Z_{L\times L}(T)^3} {Z_{3L\times L}(T)}
  = \frac{\left(\includetikz{01}\right)^3}{\includetikz{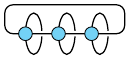}},
\end{equation}
\begin{equation}
  X'_{3\times 1}(T, L)
  \equiv \frac{\left(\includetikz{01}\right)^3}{\includetikz{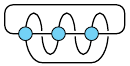}},
\end{equation}
where the prime in $X'_{3\times 1}$ indicates that the denominator corresponds to a skewed system like as $X_2$.
In this notation, $X_{2\times 1} = X_1$ and $X'_{2\times 1} = X_2$.
The further generalization of $X_1$ is straightforward as
\begin{equation}
  X_{m\times n}(T, L) \equiv \frac{Z_{L\times L}(T)^{mn}}
    {Z_{mL\times nL}(T)}.
  \label{eq:X_mn_definition}
\end{equation}
In the disordered phase, these quantities converge to one in the thermodynamic limit as well as $X_1$ and $X_2$.
On the other hand, the convergence value in the ordered phase is determined by the cluster size.
When the ordered phase is $Q$-fold degenerate, $X_{m\times n}(T, L)$ will converge to $Q^{mn - 1}$.
These dimensionless quantities are also expected to follow the same finite-size scaling form as Eq.~\eqref{eq:X1_scaling}.
Their universal values at criticality is predicted from CFT as
\begin{gather}
  X_{3\times 1} = \frac{Z_\text{PP}(i)^{3}}{Z_\text{PP}(i/3)} = \frac{Z_\text{PP}(i)^{3}}{Z_\text{PP}(3i)},\\
  X'_{3\times 1} = \frac{Z_\text{PP}(i)^{3}}{Z_\text{PP}\left(\frac{1+i}{3}\right)}
  = \frac{Z_\text{PP}(i)^{3}}{Z_\text{PP}\left(\frac{1+3i}{2}\right)},
\end{gather}
In the last equality, we transform $\tau$ so that it is in the standard fundamental domain of the modular group: $|\tau|\geq 1$, $-\frac{1}{2} < \text{Re}(\tau) \leq \frac{1}{2}$, and $\text{Im}(\tau) > 0$.
Such a transformation is also useful for computation of the modular invariant partition function.
The larger imaginary part of $\tau$ makes the absolute value of $q$ smaller, leading to faster convergence of the sum in Eq.~\eqref{eq:Virasoro_character}.

\section{Numerical Results}
\label{sec:numerical_results}

We consider three spin models on the square lattice, the two-dimensional Ising model, the three-state Potts model, and the four-state Potts model, which belong to different universality classes.
We perform numerical calculations using the BWTRG method~\cite{adachi2022bondweighted,morita2025multiimpurity} and compute the partition-function ratios defined in the previous section.
We compare numerical results of the partition-function ratios with the universal values predicted from CFT and analyze their size dependence and bond-dimension dependence.
The results for the anisotropic models and the honeycomb lattice models are shown in the next section and Appendix.

\subsection{Ising model}

\begin{figure}[t]
  \centering
  \includegraphics[width=\columnwidth]{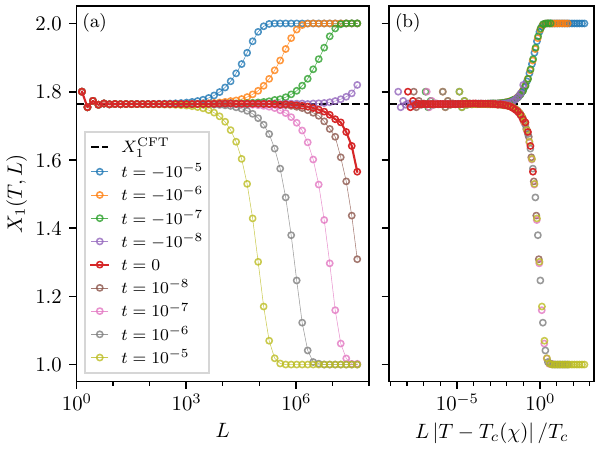}
  \caption{(a) The size dependence of $X_1$ near the critical temperature on the Ising model and (b) its FSS plot.
  The horizontal dashed line indicates the universal value predicted by CFT.
  The horizontal axis of the FSS plot is scaled by the deviation from the estimated critical temperature.
  The critical exponent related to the correlation length is taken as the exact value $\nu=1$.
  The reduced temperature is denoted by $t \equiv (T - \Tc)/\Tc$.
  Simulations were performed using BWTRG with the bond dimension $\chi=128$.}
  \label{fig:ising_X1}
\end{figure}

\begin{table}[b]
  \centering
  \caption{The universal values of $X$'s at criticality.}
  \label{tab:universal_values}
  \begin{ruledtabular}
  \begin{tabular}{lrrr}
    & Ising & $Q=3$ Potts & $Q=4$ Potts\\
    \colrule
    $X_1$            & 1.76359550 &  2.55967660 &  3.59956982 \\
    $X_2$            & 1.89631389 &  2.85152279 &  4.08334941 \\
    $X_{3\times 1}$  & 2.83994315 &  5.67899947 & 11.01929807 \\
    $X'_{3\times 1}$ & 3.52055199 &  7.88039790 & 16.06335321 \\
    $X_{4\times 1}$  & 4.34986140 & 11.56690394 & 30.30678480 \\
    $X'_{4\times 1}$ & 6.34190070 & 20.81319696 & 60.01830016 \\
  \end{tabular}
  \end{ruledtabular}
\end{table}

The Hamiltonian of the Ising model without the external magnetic field is given by
\begin{equation}
  \mathcal{H} = -J \sum_{\langle i, j \rangle} \sigma_i \sigma_j,
\end{equation}
where $\sigma_i = \pm 1$ is the Ising spin at a site $i$, and the sum is taken over all nearest-neighbor pairs.
The critical temperature is exactly known as $\Tc / J = 2 / \ln(1 + \sqrt{2}) \approx 2.269185$.

The critical phenomena of the Ising model are described by the unitary minimal CFT with the central charge $c = 1/2$.
The modular invariant partition function is given by
\begin{equation}
  Z_\text{PP}^\text{Ising}(\tau) = |\chi_{0}(q)|^2 + |\chi_{1/2}(q)|^2 + |\chi_{1/16}(q)|^2.
\end{equation}
The concrete expression of the Virasoro character $\chi_{\Delta}(\tau)$ was derived in Ref.~\cite{rocha-caridi1985vacuum}.
The universal values of $X$'s at criticality are listed in Table~\ref{tab:universal_values}.

We calculate the size dependence of $X_1$ near the critical temperature as shown in Fig.~\ref{fig:ising_X1}(a).
The bond dimension of BWTRG is set to $\chi = 128$.
It can be seen that $X_1$ takes values close to the universal value predicted by CFT over a wide range of system sizes.
The numerical data at criticality are not constant due to finite bond dimension effects.
The estimated critical temperature $\Tc(\chi)$ oscillates around the exact value as the bond dimension increases, eventually converging to the exact value~\cite{morita2025multiimpurity}.
We obtaine $\Tc(\chi) / \Tc =1 -7.93 \times 10^{-9}$ at $\chi=128$ by searching for the critical temperature using the bisection method.
Thus, the calculated value of $X_1$ at $T=\Tc$ converges to the value of the disordered phase, resulting in a downward shift.

The finite-size scaling plot of $X_1$ is shown in Fig.~\ref{fig:ising_X1}(b), where the horizontal axis is scaled by the deviation from the critical temperature.
This plot supports that $X_1$ follows the finite-size scaling form of Eq.~\eqref{eq:X1_scaling}.
We note that the horizontal axis is scaled using $(T - \Tc(\chi)) / \Tc$ instead of $t$ to account for finite bond dimension effects.
The critical exponent related to the correlation length is taken as the exact value, $\nu=1$.
In the usual finite-size scaling analysis, it is necessary to change the measured temperature points according to the system size.
However, in TRG, the system size is determined by the number of renormalization steps, so this kind of the finite-size scaling plot is useful.

\begin{figure}[t]
  \centering
  \includegraphics[width=\columnwidth]{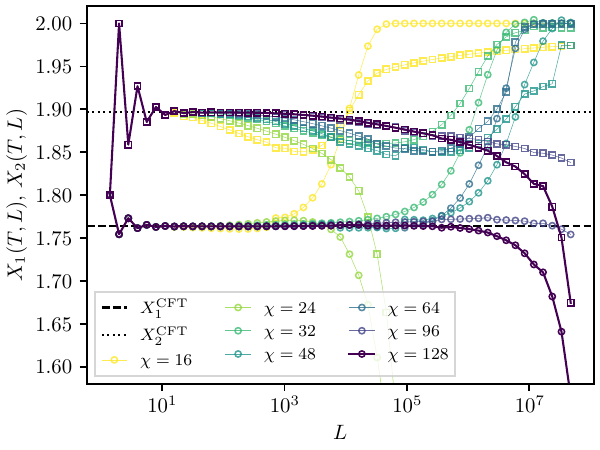}
  \caption{The bond-dimension dependence of $X_1$ (circle) and $X_2$ (square) at criticality on the Ising model.
  The dashed horizontal lines indicate the universal values predicted by CFT.}
  \label{fig:ising_X_tc}
\end{figure}

The bond-dimension dependence of $X_1$ and $X_2$ at criticality is plotted in Fig.~\ref{fig:ising_X_tc}.
As the bond dimension increases, it can be seen that the horizontal region equal to the universal values predicted by CFT expands.
Obviously, $X_2$ is more affected by the finite bond dimension than $X_1$.
Therefore, the finite-size scaling plot is not as good as that of $X_1$ in Fig.~\ref{fig:ising_X1}(b).

\begin{figure}[t]
  \centering
  \includegraphics[width=\columnwidth]{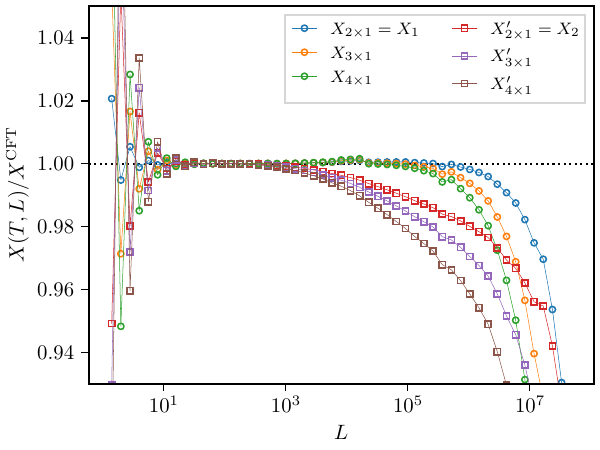}
  \caption{The size dependence of the partition-function ratios defined from larger cluster sizes at criticality on the Ising model.
  Circle symbols represent $X$'s without skew, while square symbols represent those with skew.
  Simulations were performed using BWTRG with the bond dimension $\chi=128$.}
  \label{fig:ising_X_large}
\end{figure}

We also calculate dimensionless quantities defined from larger cluster sizes using BWTRG with $\chi=128$.
Figure \ref{fig:ising_X_large} shows the deviation from the CFT universal value.
Here, circle symbols represent $X$'s without skew, while square symbols represent those with skew.
Similar to $X_1$ and $X_2$, these quantities take values close to the universal value predicted by CFT.
As a general trend, when skew is included, the deviation from the universal value becomes larger.
This may be because tensor networks on the square lattice have difficulty representing correlations in directions that are not horizontal or vertical.

\subsection{Three-state Potts model}

Next, to investigate second-order phase transitions belonging to universality classes different from the Ising model, we consider the Potts model~\cite{potts1952generalized,wu1982potts}.
The Hamiltonian of the $Q$-state Potts model without the magnetic field is given by
\begin{equation}
  \mathcal{H} = -J \sum_{\langle i, j \rangle} \delta_{\sigma_i, \sigma_j},
\end{equation}
where $\sigma_i = 1, 2, \dots, Q$ is the Potts spin at a site $i$.
The critical temperature is exactly given as $\Tc / J = 1 / \ln(1 + \sqrt{Q})$.

The critical phenomena of the three-state Potts model are described by the minimal model with central charge $c = 4/5$~\cite{dotsenko1984critical}.
Its modular invariant partition function is non-diagonal and given by
\begin{multline}
    Z_\text{PP}^\text{3-Potts}(\tau) =
    |\chi_{0}(q) + \chi_{3}(q)|^2 \\
    + |\chi_{2/5}(q) + \chi_{7/5}(q)|^2 \\
    + 2|\chi_{1/15}(q)|^2
    + 2|\chi_{2/3}(q)|^2.
\end{multline}
The universal values of $X$'s of the three-state are listed in Table~\ref{tab:universal_values}.

\begin{figure}[t]
  \centering
  \includegraphics[width=\columnwidth]{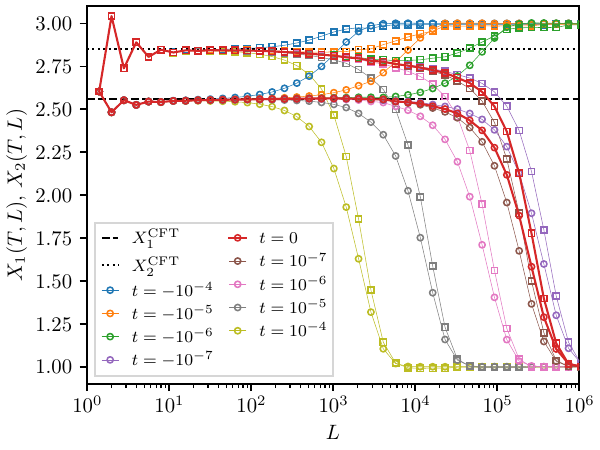}

  \caption{The size dependence of $X_1$ (circle) and $X_2$ (square) near the critical temperature on the three-state Potts model. 
  The red symbols represent the data at criticality.
  The horizontal dashed (dotted) line indicates the universal value of $X_1$ ($X_2$) predicted by CFT.
  Simulations were performed using BWTRG with the bond dimension $\chi=144$.}
  \label{fig:3potts}
\end{figure}
Figure~\ref{fig:3potts} shows the size dependence of $X_1$ and $X_2$ on the three-state Potts model, which is similar to that of the Ising model.
The values of both $X_1$ and $X_2$ near the critical temperature are close to the universal values predicted by CFT.
The data at criticality converge to unity in the thermodynamic limit, indicating that the system belongs to the disordered phase.
This incorrect behavior is because the estimated transition temperature from BWTRG with $\chi=144$ is lower than the true critical temperature.
We obtain $\Tc(\chi) / \Tc = 1 - 2.76 \times 10^{-7}$ at $\chi=144$ by searching for the critical temperature using the bisection method.
From the FSS analysis of the $X_1$ data in Fig.~\ref{fig:3potts}, we obtain the critical exponent $1/\nu = 1.23$.
The exact solution is $1/\nu = 6/5 = 1.2$, and the accuracy is improved compared to the value $1/\nu\sim 1.3$ obtained from the FSS analysis of the Binder parameter using HOTRG~\cite{morita2019calculation}.
This improvement is largely attributable to the fact that calculations with larger bond dimensions are possible by suppressing the computational cost with BWTRG.

\subsection{Four-state Potts model}

Finally, we consider the four-state Potts model.
Its critical phenomena corresponds to the $\mathbb{Z}_2$ orbifold of the compact boson CFT, which has the central charge $c=1$.
The modular invariant partition function is derived as the special case of the Ashkin-Teller model~\cite{yang1987modular,baake1987operator1,baake1987operator2,difrancesco1987relations}.
(The explicit form, being too long, is not shown here.)
Unlike the Ising model and the three-state Potts model, it has an infinite number of primary fields.
The universal values of $X$'s are listed in Table~\ref{tab:universal_values}.
The universal values of $X_2$ and $X'_{3\times 1}$ are larger than these convergence values in the ordered phase.
Therefore, $X_2(T, L)$ and $X'_{3\times 1}$ will behave non-monotonically with respect to temperature.
Such non-monotonic behavior of dimensionless quantities has also been observed in the Binder parameter~\cite{watanabe2023nonmonotonic}.

\begin{figure}[t]
  \centering
  \includegraphics[width=\columnwidth]{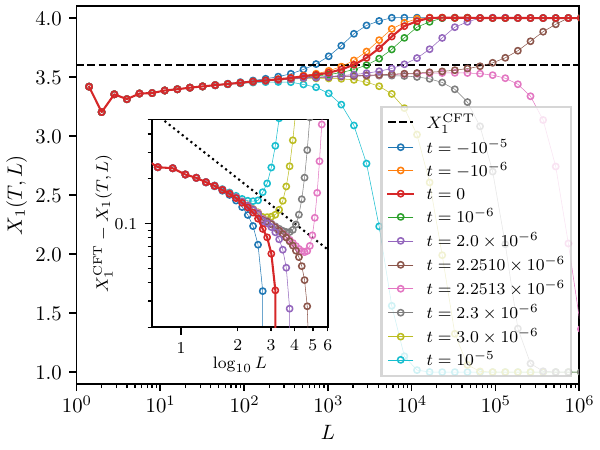}
  \caption{The size dependence of $X_1$ near the critical temperature on the four-state Potts model.
    The red symbols represent the data at criticality.
    The horizontal dashed line indicates the universal value of $X_1$ predicted by CFT.
    The inset shows the deviation from the universal value as a function of $\log_{10} L$.
    The dotted line is proportional to $(\log L)^{-1}$.
    Simulations were performed using BWTRG with the bond dimension $\chi=144$.
  }
  \label{fig:4potts}
\end{figure}

Figure~\ref{fig:4potts} shows the size dependence of $X_1$ on the four-state Potts model.
Unlike the other two models, there is no region where it agrees with the universal value from CFT.
Instead, $X_1$ seems to approach the universal value in the thermodynamic limit.
This is considered to be due to the effect of logarithmic corrections known in the four-state Potts model~\cite{cardy1980scaling,salas1997logarithmic}.
A ratio of the partition functions at criticality in the finite-size system is expected to behave as
\begin{equation}
  X(\Tc, L) \simeq g\left(\frac{1}{\log L}\right).
  \label{eq:log_correction}
\end{equation}
The inset of Fig.~\ref{fig:4potts} shows the deviation of $X_1$ from the universal value as a function of $\log_{10} L$.
The deviation decreases inversely proportional to $\log L$ near the estimated critical temperature, which is $\Tc(\chi) / \Tc = 1 + 2.251 \times 10^{-6}$ with $\chi=144$.
This result is consistent with Eq.~\eqref{eq:log_correction}.
For $X_2$, the finite bond dimension effects are more significant than those of $X_1$.
In simulations with feasible bond dimensions, the decrease of $X_2$ due to finite bond dimension appears before approaching to the universal value,
making it difficult to clearly confirm the logarithmic correction behavior.

\section{Anisotropic Cases}
\label{sec:anisotropic}

So far, we have considered isotropic models, but here we discuss anisotropic models where the correlation lengths in the horizontal and vertical directions are different.
We assume weak anisotropy, where the horizontal correlation length $\xi_x$ and the vertical correlation length $\xi_y$ are different, but the critical behavior is the same as $t^{-\nu}$.
Since CFT has rotational symmetry, to obtain an effective CFT from a lattice model, it is necessary to perform different scale transformations corresponding to the correlation lengths in the vertical and horizontal directions.
Therefore, a square system of size $L \times L$ effectively has an aspect ratio of $\xi_x^{-1}:\xi_y^{-1}$ in CFT.
That is, the modular parameter changes from $\tau = i$ to $\tau = i\xi_x/\xi_y$.

The above discussion modifies the universal value of the partition-function ratio at criticality as
\begin{equation}
  X_1 = \frac{Z_\text{PP}\left(i\xi_x / \xi_y \right)^2}
  {Z_\text{PP}\left(i\xi_x / 2\xi_y\right)}
  = \frac{Z_\text{PP}\left(i\xi_y / \xi_x\right)^2}
  {Z_\text{PP}\left(2i\xi_y / \xi_x\right)}.
  \label{eq:X1_anisotropic}
\end{equation}
We assume that, in the denominator of the partition-function ratio (Eq.~\eqref{eq:X1_definition}), the horizontal direction has twice the length.
Extension to $X_2$ and others is straightforward.

In TRG calculations, it is necessary to consider the orientation of tensor networks.
After an odd number of TRG steps, the tensor network is tilted at 45 degrees, and the coarse-grained tensor corresponds to a square region rotated by 45 degrees.
By performing scale transformations to make it have isotropic correlations, this is transformed into a rhombus with diagonals having $\xi_x^{-1}$ and $\xi_y^{-1}$.
Therefore, when using the 45-degree tilted tensor obtained after an odd number of TRG steps~\eqref{eq:X1_tensor}, the universal value of $X_1$ is given by
\begin{equation}
  \tilde{X}_1 = \frac{Z_\text{PP}\left(e^{2i\theta}\right)^2}
  {Z_\text{PP}\left(e^{2i\theta}/2\right)}
  \label{eq:X1_anisotropic_45deg}
\end{equation}
where $\theta \equiv \arctan\left(\xi_y / \xi_x\right)$.

\begin{figure}[t]
  \centering
  \includegraphics[width=\columnwidth]{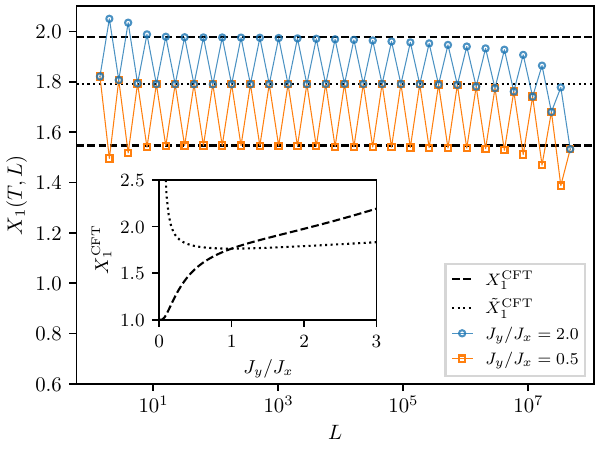}
  \caption{The size dependence of $X_1$ at criticality on the anisotropic Ising model for $J_y/J_x=2.0$ (circle) and $0.5$ (square).
  The dashed (dotted) horizontal lines indicate the universal values for the untilted (45-degree tilted) tensor networks predicted by CFT.
  Simulations were performed using BWTRG with the bond dimension $\chi=128$.
  The inset shows the universal values of $X_1$ and $\tilde{X}_1$ as a function of $J_y/J_x$.}
  \label{fig:anisotropic_ising}
\end{figure}

We numerically verify this argument in the anisotropic Ising model on the square lattice.
Let the nearest-neighbor coupling constants in the horizontal and vertical directions be $J_x$ and $J_y$, respectively.
Even with anisotropy, the two-dimensional Ising model has an exact solution and belongs to the same universality class as the isotropic case.
The critical temperature $\Tc$ is given by the relation
\begin{equation}
  \sinh\left(2J_x/\Tc\right)
  \sinh\left(2J_y/\Tc\right) = 1.
\end{equation}
The horizontal correlation length at temperature $T$ is given by
\begin{equation}
  \xi_x^{-1} = \begin{cases}
    4 J_y / T - 2 \ln \coth\left(J_x / T\right) & (T < \Tc), \\
    \ln \coth\left(J_x / T\right) - 2 J_y / T & (T > \Tc),
  \end{cases}
\end{equation}
and the vertical correlation length is given by the same expression with $x$ and $y$ interchanged~\cite{wu1966theory}.
The correlation-length ratio in the limit of $T\rightarrow \Tc$ is
\begin{equation}
  \lim_{T\rightarrow\Tc}\frac{\xi_y}{\xi_x} = \sinh\left(\frac{2J_y}{\Tc}\right)
  =\sqrt{\frac{\sinh(2J_y/\Tc)}{\sinh(2J_x/\Tc)}}.
\end{equation}
Therefore, the universal value of the partition-function ratio predicted from CFT can be calculated as a function of $J_y/J_x$.
The inset of Fig.~\ref{fig:anisotropic_ising} shows the universal value of $X_1$ plotted as a function of $J_y/J_x$.
The universal value for the untilted tensor network increases monotonically with $J_y/J_x$.
On the other hand, the universal value for the 45-degree tilted tensor network takes a minimum value at $J_y/J_x=1$, i.e., the isotropic case.

Figure~\ref{fig:anisotropic_ising} shows the size dependence of $X_1$ at criticality for $J_y/J_x=2.0$ and $0.5$ calculated using BWTRG with $\chi=128$.
It can be confirmed that they agree with the universal value predicted by Eq.~\eqref{eq:X1_anisotropic} and Eq.~\eqref{eq:X1_anisotropic_45deg}.
The reason why the partition-function ratios for the 45-degree tilted tensor network are equal for $J_y/J_x=2.0$ and $0.5$ is that they are transformed into each other through swapping the $x$-axis and $y$-axis.
Compared to the isotropic case, the values at criticality deviate from the universal value for small systems.
This is considered to be due to insufficient representation capability in the direction where the correlation length becomes longer due to anisotropy.

\section{Summary and Discussion}
\label{sec:summary}

We have investigated the behavior of dimensionless quantities defined as ratios of the partition functions.
These quantities are easier to compute using TRG than the Binder parameter and follow the same finite-size scaling form as other dimensionless quantities including the Binder parameter.
At criticality, they take the universal values that can be predicted from CFT.
We performed numerical simulation using the BWTRG method and confirmed these properties for the two-dimensional Ising model and the three-state Potts model.
In the four-state Potts model, we found that the ratio of the partition functions does not take a constant value with respect to the system size.
The deviation from the universal value decreases inversely proportional to the logarithm of the system size, which is consistent with the logarithmic corrections known in the four-state Potts model.
Direct observation of logarithmic corrections is usually a difficult task, and our results demonstrate the usefulness of calculations using tensor networks.

We have also considered ratios of the partition functions defined from larger cluster sizes in addition to $X_1$ and $X_2$ proposed by Gu and Wen, and confirmed that similar properties hold.
In particular, the universal value of the partition-function ratio at the critical point can be predicted from the CFT, even for different cluster sizes.
We found that the relative error from the universal value increases as the cluster size becomes larger.
To accurately calculate the partition function for large clusters, it is necessary to correctly represent long-range correlations.
Therefore, for large cluster sizes, the bond dimension is insufficient to achieve a certain level of accuracy.
We also found that when skew is included, the deviation from the universal value tends to be larger.
It is because a tensor network on square lattices has difficulty representing correlations in diagonal directions.

In anisotropic case, we have discussed that the universal value of the partition-function ratio depends on the ratio of correlation lengths.
We confirmed that the universal value predicted from CFT agree with the numerical results in the anisotropic Ising model, since the exact solution allows us to calculate the ratio of correlation lengths from the ratio of interactions.
In general models, the ratio of correlation lengths cannot be calculated exactly, but it is possible to estimate the ratio of correlation lengths from the universal value of the partition-function ratio.
It is known that the accuracy of finite-size scaling analysis with minimal computational cost improves by using the aspect ratio of the system according to the correlation lengths~\cite{matsumoto2001groundstate,yasuda2013monte}, and our method is expected to be a useful technique for that purpose.

In this paper, we have assumed a tensor network structure on a uniform square lattice.
That is, the tensors at each site are all identical.
In the case of a square lattice with multiple different tensors, as seen in TRG with entanglement filtering technique~\cite{gu2009tensorentanglementfiltering,evenbly2015tensor,yang2017loop,harada2018entanglement,hauru2018renormalization,homma2024nuclear}, the definition in Eq.~\eqref{eq:X1_tensor} does not work and it should be extended to use a larger cluster as the basic unit.
For instance, $X_1$ can be defined using a $2\times 2$ cluster as the unit,
\begin{equation}
  X_1^{(2\times 2)}(T, L)
  \equiv \frac{Z_{2L\times 2L}(T)^2}{Z_{4L\times 2L}(T)}
  = \frac{\left(\includetikz{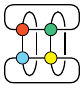}\right)^2}{\includetikz{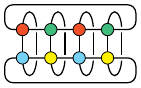}},
\end{equation}
whose computational cost is proportional to $\chi^6$.
Ignoring the difference in computational cost, this definition generally provides better accuracy than the original definition in Eq.~\eqref{eq:X1_tensor},
because its value is determined from the eigenvalue of the transfer matrix with width $2L$.
In general, longer length of the shortest side is expected to capture larger scaling dimensions and improves accuracy.

So far, we have considered only the partition-function ratios where the numerator is the partition function for a square region.
More generally, it is also possible to define dimensionless quantities in which the numerator is the partition function for a rectangular region like as ${Z_{mL\times L}(T)^n}/{Z_{mL\times nL}(T)}$.
As mentioned above, this quantity will have the precision of the transfer matrix with width $L$.
Therefore, generalizations in this direction are thought to offer little benefit.

Comparing with the method of obtaining scaling dimensions from the eigenvalues of the transfer matrix, the information about critical phenomena obtained from the finite-size scaling of the ratio of partition functions is limited.
As is clear from the definition in Eq.~\eqref{eq:X1_definition}, the ratio of partition functions can be calculated from the eigenvalues of the transfer matrix, and its behavior at criticality is dominated by small scaling dimensions.
Therefore, the system size at which the ratio of partition functions begins to deviate from the universal value almost coincides with the system size at which the accuracy of calculating scaling dimensions becomes low.
Calculating scaling dimensions becomes difficult even slightly away from the estimated critical temperature.
However, the partition-function ratio can be calculated at any temperature, which is the advantage of this method.

It is known that redundant short-range correlations make the accurate calculation of large scaling dimensions difficult.
The BWTRG method used in this paper appears to mitigate the short-range correlation problem compared to TRG, but it still remains.
It should be possible to robustly compute the universal value using a method with entanglement filtering~\cite{gu2009tensorentanglementfiltering,evenbly2015tensor,yang2017loop,harada2018entanglement,hauru2018renormalization,homma2024nuclear}, but this remains a topic for future research.

\begin{acknowledgments}
  The present work was supported by JSPS KAKENHI Grants No.~20K03780, No.~23H01092 and No.~23H03818.
  S.M. is supported by the Center of Innovations for Sustainable Quantum AI (JST Grant Number JPMJPF2221).
  The computation in this work was done using the facilities of the Supercomputer Center, the Institute for Solid State Physics, the University of Tokyo (ISSPkyodo-SC-2024-Ba-0019, 2025-Ba-0040).
\end{acknowledgments}

\subsection*{DATA AVAILABILITY STATEMENT}
The data that support the findings of this study are openly available in the ISSP Data Repository~\cite{isspdatarepo}.

\appendix*
\section{Honeycomb Lattice}\label{sec:honeycomb}

In this appendix, we consider the ratios of partition functions on the honeycomb lattice.
As a natural extension of $X_1$ to a tensor network on the honeycomb lattice, we define the following ratio of partition functions:
\begin{equation}
  X_{\text{H}, 1}(T, L)
  \equiv \frac{\left(\includetikz{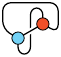}\right)^2}{\includetikz{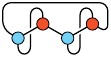}}.
\end{equation}
Its convergence values in the ordered phase and disordered phase are the same as those of the square lattice.
In the disordered phase, it converges to one, and in the ordered phase with $Q$-fold degeneracy, it converges to $Q$.
Its finite-size scaling is also expected to take the same form.
Because of the symmetry of the honeycomb lattice, a similar extension of $X_2$ to the honeycomb lattice is equivalent to $X_{\text{H}, 1}$.

\begin{table}[b]
  \centering
  \caption{The universal values of $X_{\text{H}, 1}$ and $X_{\text{H}, 3}$ at criticality.}
  \label{tab:universal_values_honeycomb}
  \begin{ruledtabular}
  \begin{tabular}{lrrr}
    & Ising & $Q=3$ Potts & $Q=4$ Potts\\
    \colrule
    $X_{\text{H}, 1}$ & 1.80609893 &  2.65270688 &  3.74732258 \\
    $X_{\text{H}, 3}$ & 3.57165236 &  8.04919013 & 16.46486997
  \end{tabular}
  \end{ruledtabular}
\end{table}

\begin{figure}[t]
  \centering
  \includegraphics[width=\columnwidth]{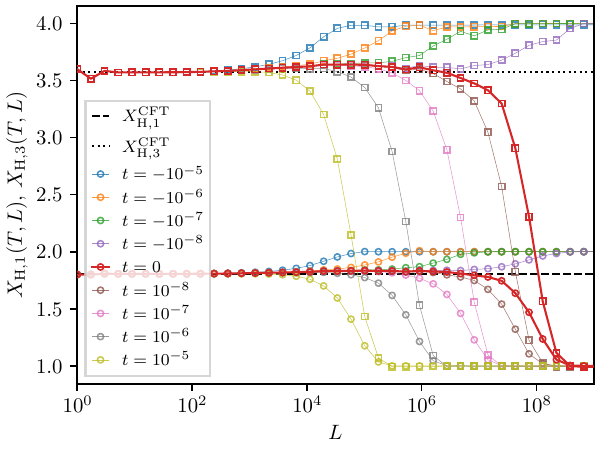}
  \caption{The size dependence of $X_{\text{H}, 1}$ (circle) and $X_{\text{H}, 3}$ (square) in the Ising model on the honeycomb lattice near criticality.
  The red symbols represent the data at criticality.
  The horizontal dashed (dotted) line indicates the universal values of $X_{\text{H}, 1}$ ($X_{\text{H}, 3}$) predicted by CFT.
  Simulations were performed using BWTRG with the bond dimension $\chi=128$.}
  \label{fig:ising_hex}
\end{figure}

The universal value of $X_{\text{H}, 1}$ at criticality is different from that of $X_1$.
This is because the shape of the unit cell of the honeycomb lattice is a rhombus formed by joining two equilateral triangles, not a square.
The universal value of $X_{\text{H}, 1}$ is predicted from CFT using the modular invariant partition function as
\begin{equation}
  X_{\text{H}, 1} =
  \frac{Z_\text{PP}\left(\frac{1+\sqrt{3}i}{2}\right)^2}
  {Z_\text{PP}\left(\frac{1+\sqrt{3}i}{4}\right)}
  = \frac{Z_\text{PP}\left(\frac{1+\sqrt{3}i}{2}\right)^2}
  {Z_\text{PP}\left(\sqrt{3}i\right)}.
\end{equation}
The universal values of $X$'s of the Ising model, the three-state Potts model, and the four-state Potts model on the honeycomb lattice are listed in Table~\ref{tab:universal_values_honeycomb}.

We can define another ratio of partition functions specific to the honeycomb lattice as
\begin{equation}
  X_{\text{H}, 3}(T, L)
  = \frac{\left(\includetikz{10}\right)^3}{\includetikz{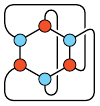}}.
\end{equation}
It converges to one in the disordered phase, while it converges to $Q^2$ in the ordered phase with $Q$-fold degeneracy.
The denominator of $X_{\text{H}, 3}(T, L)$ is equivalent to the partition function of a rhombus system scaled by $\sqrt{3}$.
In other words, applying one step of TRG to the denominator results in the same diagram at the numerator.
Thus, the modular parameter is common to both the denominator and the numerator.
Therefore, the universal value of $X_{\text{H}, 3}$ at criticality is predicted from CFT as
\begin{equation}
  X_{\text{H}, 3} = Z_\text{PP}\left(\frac{1+\sqrt{3}i}{2}\right)^{2}.
\end{equation}

As a numerical demonstration, we calculate $X_{\text{H}, 1}$ and $X_{\text{H}, 3}$ in the Ising model on the honeycomb lattice using BWTRG with the bond dimension $\chi=128$.
The results are shown in Fig.~\ref{fig:ising_hex}.
Similar to the square lattice case, both quantities near the critical temperature agree with the prediction of CFT.

\newpage
\bibliography{ratio_Z,datarepo}

\end{document}